\documentclass[twocolumn]{aa}
\usepackage{graphicx}
\usepackage{ifthen}
\usepackage{natbib}
\usepackage{txfonts}

\def\CompactFigs{0}

\def\ltsima{$\; \buildrel < \over \sim \;$}
\def\simlt{\lower.5ex\hbox{\ltsima}}
\def\gtsima{$\; \buildrel > \over \sim \;$}
\def\simgt{\lower.5ex\hbox{\gtsima}}
\def\arcsec{\mathop{\rm arcsec}\nolimits} 

\def\mi{M_I^{TRGB}}
\def\mj{M_J^{TRGB}}
\def\mh{M_H^{TRGB}}
\def\mk{M_K^{TRGB}}
\begin{document}
   \title{The calibration of the RGB Tip as a Standard Candle}

   \subtitle{{Extension to Near Infrared colors and higher metallicity}
   \thanks{Based on observations collected at ESO, La Silla, Chile. }}

   \author{M. Bellazzini\inst{1}, F.R. Ferraro\inst{2}, A. Sollima\inst{1,2},
   E. Pancino\inst{1}, L. Origlia\inst{1}}

   \offprints{M. Bellazzini}

\institute{INAF - Osservatorio Astronomico di Bologna, Via Ranzani 1, 40127, 
          Bologna, Italy                
              \email{bellazzini@bo.astro.it}
	 \and 
	 Dipartimento di Astronomia, Universit\`a di Bologna, 
	 Via Ranzani 1, 40127, Bologna, Italy
	 \email{ferraro@bo.astro.it}
	      }

   \date{DRAFT 0.0 - \today}

\abstract{ 
New empirical calibrations of the Red Giant Branch Tip in the I,J,H and K bands 
based on two fundamental pillars, namely $\omega$~Centauri and 47 Tucanae, 
have been obtained by using a large optical and near infrared photometric 
database. 

Our best estimates give  
$\mi = -4.05 \pm 0.12$, $\mj = -5.20 \pm 0.16$, $\mh = -5.94 \pm 0.18$ 
and $\mk = -6.04 \pm 0.16$ at $[M/H]\simeq-1.5$ ($\omega$~Cen) and 
$\mi = -3.91 \pm 0.13$, $\mj = -5.47 \pm 0.25$, $\mh = -6.35 \pm 0.30$ 
and $\mk = -6.56 \pm 0.20$ at $[M/H]\simeq-0.6$ (47~Tuc).  

With these new empirical calibrations we also provide robust relations of the
TRGB magnitude in I, J, H and K bands as a function of the global metallicity.
It has also been shown that our calibrations self-consistently provide a
distance modulus of the Large Magellanic Cloud in good agreement with the
standard value ($(m-M)_0\simeq 18.50$)).

}

   \maketitle
%

\section{Introduction}

The use of Tip of the Red Giant Branch (TRGB) as a standard candle is now a
mature and widely used technique to estimate the distance to
galaxies of any morphological type 
\cite[see][for a detailed description of the method, recent reviews and
applications]{lfm93,mf95,mf98,walk}. The underlying physics is well understood
\citep{mf98,scw} and the observational procedure is operationally well defined
\citep{mf95}. The key observable is the sharp cut-off occurring at the bright
end of the RGB Luminosity Function that can be easily detected with the
application of an edge-detector filter \citep[Sobel filter,][]{mf95,smf96} or by
fitting the LF with a proper modeling function \citep{mendez}. The necessary
condition for a safe application of the technique is that the observed
RGB Luminosity Function should be well populated, with more than $\sim 100$ 
stars within 1 mag from the TRGB \citep{mf95,draco}. 

The RGB develops to its full extension in stellar populations having 
age \gtsima 1-2 Gyr, therefore
the natural local systems with which the method may be calibrated are Galactic
globular clusters \citep{da90,f00}. However, any calibration based on globulars
must rely on the RR Lyrae (or Horizontal Branch) distance scale, whose
zero-point is still quite uncertain \citep{carla,walk}. Moreover the estimate
of the TRGB in globular clusters is plagued by the low number of RGB stars
available in each cluster, implying a considerable statistical uncertainty 
and forcing the adoption of a different operational definition of the observable
\cite[e.g., the brightest non-variable RGB star instead of the LF cut-off, 
see][and reference therein, for details and discussion]{da90,f00,tip}. 

To overcome these drawbacks
\citet[][hereafter Pap-I]{tip} used a large photometric database of the globular
cluster $\omega$~Centauri (the most luminous in the whole Galaxy, therefore the
one having the largest number of RGB stars) and the distance estimate to this
cluster by 
\citet[][based on the double-lined detached eclipsing binary OGLE 17]{ogle}, to
provide an accurate zero point to the TRGB calibration in the Cousins' I
passband, independent of the RR Lyrae distance scale. The detection of the 
cut-off in the LF of the upper RGB was clean and supported by a large sample 
comprising more that 180 stars in the brightest 1 mag of the RGB, e.g. fully
satisfying all the requirements for a safe estimate of the TRGB magnitude as
defined by \citet{mf95} and \citet{draco} by means of numerical experiments.
Furthermore the use of an essentially geometrical distance estimate to the
considered cluster to calibrate the TRGB makes it, in practice, a {\em primary}
standard candle. The calibration has been already applied by several authors in
different contexts \citep{tosi,apella,draco,alves,alves2}.

In the present paper we shortly re-discuss the analysis of Pap-I at the light of
some new results appeared in the literature and we extend the calibration to the
Near Infrared (NIR) passbands with a statistically robust estimate of the
absolute magnitude of the TRGB of $\omega$~Cen in the J,H and K bands based on the 
large NIR photometric
database assembled by \citet{antonio}, complemented with 2MASS \citep{cutri} 
data. While the metallicity dependence of the magnitude of the TRGB
is minimized in the Cousins' I band, the most suitable for distance estimates,
the calibration of the standard candle in the NIR may prove very useful since
(a) it may allow a safe application also in the cases of very high interstellar
extinction, and (b) it will provide a natural tool for future studies performed
with powerful telescopes optimized for infrared observations, as, for example, 
the James Webb Telescope that will push the application limit of the TRGB
technique to much larger distances. Nevertheless, it has to be recalled
that a safe application of the NIR calibrations requires a good estimate of the
metallicity of the considered stellar system, because of the strong dependence
of the TRGB brightness in these passbands (K and H, in particular).

To check our calibration also at different metallicity regimes (with respect
to $\omega$~Cen) we obtained B,V,I photometry of more that 100000 stars in a
$\sim 0.5\times 0.5$ deg$^2$ field centered on the metal rich globular cluster
47~Tucanae. With this new large sample, complemented with NIR photometry from
\citet{paolo} and 2MASS, we estimate the TRGB magnitude in I,J,H and K bands also in
this cluster. 

The plan of the paper is the following: in Sect.~2 we shortly re-discuss the
calibration of Pap-I and we consider the possibility of extension of the 
calibration to other clusters (i.e. metallicity); 
in Sect.~3 and 4 we present our new estimates of the absolute
magnitude of the TRGB in the various considered passbands for $\omega$~Cen and 
47~Tuc, respectively; in Sect.~5 we present the new calibrations of the TRGB in
I, J, H and K bands as a function of the global metallicity 
([M/H], see \citet{scs93}) and we compare our 
derived distance scale with the ``standard'' distance
modulus of the Large Magellanic Cloud \cite[$(m-M)_0=18.50$, see][]{alves2}.
Finally, the main results of the paper are summarized in Sect.~6.

\section{Refinements and extensions of the TRGB calibration}

The estimate of the apparent I magnitude of the TRGB of $\omega$~Cen derived in
Pap-I is based on a very large sample \cite[more than 220000 stars,
see][]{panc}, the final observational uncertainties have been accurately
estimated and can be considered as small as possible given the present
technological resources. 

On the other hand it was suggested in Pap-I that the
uncertainties on the other two
ingredients of the obtained zero-point of the calibration, e.g. distance and
reddening, may be significantly reduced with better data. For instance,
according to \citet{ogle}, a better coverage of the velocity curve of 
OGLE-17 and/or better estimate of the temperature of the members of the binary
should significantly enhance the accuracy of the distance estimate. 
While no new result become available since the publication of Pap-I on this
side, there are interesting news concerning the reddening. In a recent review
\citet{lub} pointed out that while the usually assumed E(B-V) values range from
0.09 to 0.16 mag, the actual reddening of $\omega$~Cen has been very accurately
determined by a number of authors in the '70 and '80. All the best independent
methods consistently give $E(B-V)=0.11 \pm 0.01$, a more accurate and robust
estimate than that adopted by \citet{ogle} and in Pap-I 
(e.g., $E(B-V)=0.13 \pm 0.02$). From now on we will follow the indications of
\citet{lub}, adopting $E(B-V)=0.11 \pm 0.01$.

First of all we repeated the analysis of \citet{ogle} 
with the newly assumed reddening, to check the effect on the 
{\em true distance modulus} of $\omega$~Cen. The new estimate of the linear 
distance to the cluster is $D=5500 \pm 300$ pc (instead of $5350 \pm 300$).
However, since the extinction enters in two of the equations needed to
obtain the apparent distance modulus in V \cite[eq. 2 and 4 in][]{ogle}, the
effect of the new assumption cancels out and the final $(m-M)_V$ is unchanged,
$(m-M)_V=14.04 \pm 0.11$. The {\em true} distance modulus is $(m-M)_0=13.70 \pm
0.11$ (instead of $(m-M)_0=13.64 \pm 0.11$), but the effect cancels out again in
the determination of the absolute I magnitude of the TRGB $\mi = -4.05 \pm
0.12$, just one hundredth of magnitude brighter than what found in Pap-I.

The distance modulus
derived above is in excellent agreement with the most recent estimates 
obtained with different methods, e.g. the HB semi-empirical fitting by F99 and
\citet{rey}, the theoretical Period-Luminosity-Metallicity relation used by
\citet{caputo} and the RR Lyrae distance scale obtained by \citet{g03} from
the more recent and finest application of the Main Sequence fitting of galactic
globulars to local subdwarfs whose distance has been measured with the Hipparcos
satellite \cite[see also][]{olech}. 
Hence the fundamental pillars of our calibration seem to have gained
further support from the advancements made in this field since the publication
of Pap-I.

\subsection{Metallicity, age and contamination issues}

It is well known that $\omega$ Cen is not a Simple Stellar Population
\cite[SSP][]{rf88} but instead it host stars of different metallicities and ages
\cite[see][and references therein]{ndc,n96,sk96,hilker,hughes,panc,panct}.
In Pap-I we shortly discuss the main reasons that lead us to consider
this stellar system as a suitable pillar of the TRGB calibration in spite of
these possible shortcomings. Here, following an indication of the referee, 
we illustrate these arguments in deeper
detail, considering the potential problems associated with the presence of a
metallicity and age spread and with foreground contamination by stars of
Galactic disc.

Several photometric and spectroscopic studies have ascertained that $\omega$ Cen
hosts stars of widely different chemical abundance, from $[Fe/H]\simeq -2.0$ to
$[Fe/H]\simeq -0.5$ \citep{ndc,panc02,livia}. However, the results of the 
largest spectroscopic surveys clearly indicate that the metallicity 
distribution of  $\omega$ Cen is characterized by an obvious main peak around
$[Fe/H]\simeq -1.7$ (corresponding to $[M/H]\simeq -1.45$, see Sect.~5, below).
From the spectroscopic metallicity estimates for 144 RGB stars by \citet{sk96}
it can be concluded that more than 80\% of the cluster stars are enclosed in the
narrow range $[Fe/H]= -1.7\pm 0.2$. The same conclusion can be drawn from the
survey by \citet{n96} of 517 RGB stars and from the large set of 
metallicities obtained from Str\"omgren photometry by \citet{hilker}. Hence, it
can be concluded that the stellar content of $\omega$ Cen is largely
dominated by a stellar population spanning a narrow range in metallicity (and
presumably age) and it can be safely considered as an approximate SSP in the 
present context.
In Sect.~3.1, in the process of detection of the TRGB in the NIR passbands, 
we will select RGB stars in color to avoid possible contaminations by stars
not associated with the main peak of the metallicity distribution.

The weak sensitivity of the TRGB luminosity to age, for ages larger than $\sim
4$ Gyr, constitutes the original theoretical basis of its use as a standard
candle. At a given chemical composition the TRGB level is determined by the mass
of the stellar core at the Helium flash, that is fairly constant over a large
part of the low-mass star range \cite[see][and references therein]{scw,mf98}.
Note that this is the same basic reason that allow the use of RR Lyrae and/or
Horizontal Branch stars as standard candles. Either using the models adopted 
by \citet{sc98}, or the analytical formulae provided by \citet{rood}, or those
by \citet{buz,buz2}, we easily recover the well known result that the
bolometric magnitude ($M_{bol}$) of the TRGB decrease by less than 0.1 mag 
while varying the
age from 14 Gyr to 4 Gyr (our analysis is limited to stellar populations with
less-than-solar metallicity). We considered also the impact of bolometric
corrections (BC), adopting the prescriptions by \citet{da90} for the Cousins' I and
those by \citet{paolo} for J and K. We found that the difference in
bolometric correction at the TRGB due to age differences in the considered range
is (a) less than $ 0.05$ mag in I and J and less than $0.2$ mag in K, and (b) 
in J and K (and so presumably also in H) it goes in the opposite direction 
with respect to the
behavior of $M_{bol}$. It turns out that, at any metallicity, the level of the
TRGB in the J passband is the less sensitive to age variations and that, in any
of the considered passbands, the TRGB level varies less than $0.06$ mag for an
age difference of 6 Gyr, and less than $0.12$ mag for and age difference as
large as 10 Gyr. 

Even more relevant, in the present contest, is the recent analysis by
\citet{barker}. These authors studied the dependence of the TRGB level of
composite stellar populations on the adopted star formation history (SFH)
by means of extensive experiments with
synthetic CMDs. They conclude that ``...the TRGB-distance method is insensitive
to star formation history except for large bursts between ages of about 1 and 2
Gyr''. The maximum age spread suggested by \citet{hilker,hughes} for $\omega$
Cen is $\le 6$ Gyr and the stars younger than the metal poor bulk of the
cluster stars are a small fraction of its whole population. Hence we should
conclude that the observed age spread is not a concern for the use of $\omega$
Cen as a fundamental pillar of our TRGB calibration. The considerations above
also demonstrate that age (or SFH) issues should not be a concern even for the
application of the method to the Large Magellanic Cloud, presented in Sect.~5.1,
below.

Finally, the relatively low Galactic latitude of $\omega$ Cen ($b\simeq 15^o$)
implies a significant contamination of the CMD by foreground/background stars
belonging to the Galactic disc. Nevertheless, the inspection of the available 
CMDs \citep{panc} suggests that disc contamination is not a serious problem at
the TRGB level for fields as the ones considered here. According to the Galactic
model by \citet{robin}, the field stars expected in the upper 1 mag
from the TRGB of $\omega$ Cen and with similar color of the cluster RGB are less
than 20 and they are nearly uniformly distributed in magnitude. Hence they
cannot have any sizeable influence on the clear detections of the cut-off of the
RGB LF of $\omega$ Cen shown in Pap-I and in Sect.~3.1, below.

\subsection{Searching for other pillars: the case of 47~Tuc}

The criterion for a safe detection of the TRGB based on $N_{\star}$
\citep{mf95,draco} may be fulfilled by collecting the photometry of
large samples of RGB stars. The task may be achieved by observing sufficiently
wide regions of the stellar system under consideration, for instance by using a
wide field instrument, as done in Pap-I. However unaffordable intrinsic
obstacles to this aim may be encountered, e.g. the considered stellar system may
not {\em have} a sufficient number of RGB stars. In a Simple Stellar Population
\cite[SSP, composed of stars of the same age and chemical composition, like a
globular cluster --][]{rf88} the number of stars in a given evolutionary phase at
any given time scales with the total luminosity according to the 
Evolutionary Flux relation \citep{rebuz}. 
For example, according to \citet{alvio}, a 15 Gyr old
SSP of solar metallicity having a total luminosity of $10^5 L_{\odot}$ should
have just 10 RGB stars within 1 mag from the TRGB, clearly insufficient for a
reliable detection of the observable. 

%
   \begin{figure*}
   \centering
   \ifthenelse{\CompactFigs=0}
   {\includegraphics[width=18cm]{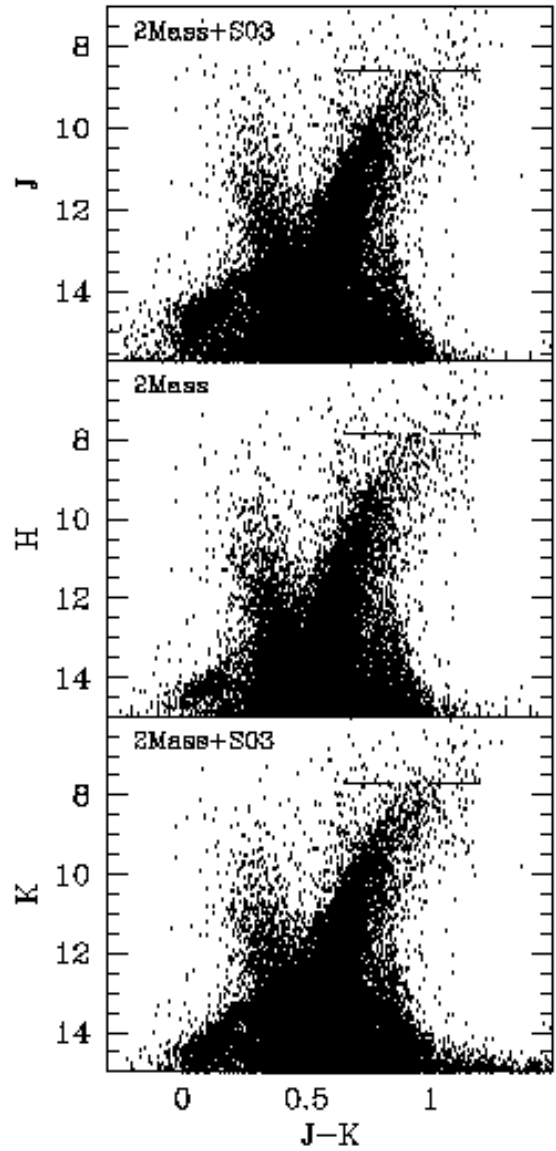}}
   \caption{NIR CMDs of $\omega$~Centauri. The position of
   the TRGB is marked by the two segments. 
    In the lower panel we also plotted the
   RGB ridge-line of the cluster M~13 ($[M/H]=-1.2$; $[Fe/H]=-1.4$) 
   taken from \citet{valenti}. 
   For the TRGB detection we selected only RGB stars to the blue of the reported
   ridge-line, in order to avoid contamination by metal-rich stars.}
    \end{figure*}
%

This basic consideration indicates that the Galactic globulars in which a safe
detection of the TRGB is possible are rare\footnote{In particular if the 
adopted observable is the cut-off in the RGB LF. This observable is much more
reliable than ``the brightest non-variable star'' when a distant stellar system
is considered, since it does not need an exact discrimination between AGB and
RGB stars (often not possible in such systems) and it is not sensitive to the 
contamination by background/foreground stars, if the degree of contamination is
not extreme \citep{mf95}.}. 

In Pap-I we found $N_{\star}\simeq
185$ for $\omega$~Cen, by far the most luminous globular cluster of the whole
Milky Way. Scaling this number with the ratio of cluster luminosity \citep[taken
from][]{harris} to the luminosity of $\omega$~Cen we find only three more
clusters for which a complete sampling of the RGB stars would achieve the
condition $N_{\star}\simgt 100$, e.g. NGC~6715, NGC~6441 and NGC~2419. Each of
them presents intrinsic problems: the CMD of NGC~6715 is strongly contaminated
by stars of the Sgr dwarf galaxy, in which is embedded \citep{sl95}; NGC~6441 is
located in the Galactic bulge and suffers from strong (and uncertain) extinction
\cite[E(B-V)=0.47,][]{harris}; NGC~2419 is one of the most distant Galactic
clusters known. For none of them, a RR Lyrae (or HB)-independent distance 
estimate is presently available. 
These factors make these cluster unsuitable as safe
pillars of the TRGB calibration, at least at the present stage.

\citet{draco}
provided a relation to correct the estimate of $I_{TRGB}$ for low-sample effects
in the range $N_{\star}\simgt 80$, thus allowing to include slightly less luminous
clusters among the potential calibrators at the expense of a larger uncertainty
in the final estimate. With this extension, the most natural candidate is
certainly $47~Tucanae$ (NGC~104) that is relatively luminous, nearby, 
has modest extinction and it is one of the most extensively studied cluster in 
the whole Galaxy. In Sect.~4 we will describe the detection of the TRGB in 
this cluster, based on a newly obtained photometric dataset including the large
majority of the 47~Tuc giants. 

\section{The TRGB of $\omega$~Cen in the J,H and K bands}

To obtain a NIR photometric database of comparable size with respect to that
used in Pap-I we merged the J,K photometry by \citet[][hereafter S03]{antonio} 
with
J,H,K photometry of a circular region centered on $\omega$~Centauri and with
1 deg radius \citep[of the order to the cluster tidal radius, see][and
references therein]{vanleuw,harris} taken from the Point Source Catalogue 
of 2MASS.
The photometry by S03 covers a field of $\simeq 13 \times 13$ arcmin$^2$ in the
central region of the cluster and provides much deeper and accurate photometry
with respect to the 2MASS catalogue. Selecting the sources for the latter we
accepted only first quality stars, with no indication of blending, corruption or
problems in the measures. These requirements make the 2MASS catalogue very 
incomplete in the inner $10 \arcmin$ where crowding is too severe for the
low spatial resolution of the survey camera. Therefore the two merged catalogues
are nicely complementary, covering different radial annuli of the cluster from
the center to the limiting radius. The tiny photometric shifts described by S03
have been applied to the 2MASS catalogue in order to match the photometric
system of S03. 
The comparison between the stars
in common between the two catalogue suggest that the final photometries are in
the same photometric system within $\pm 0.03 $ mag in J and K and the star to
star scatter at any magnitude is $\simlt 0.1$ mag. 
These figures have to be considered
as upper limits since the comparison is necessarily performed in the small
overlapping region in which the crowding may still sensibly affect the
accuracy of 2MASS photometry of individual stars. 
In the final merged catalogue we used the 2MASS dataset to complement 
the S03 catalogue both in radial extension (all the stars outside the S03 field) 
and in photometry (H band measurements not included in the S03 catalogue).
%
   \begin{figure*}
   \centering
   \ifthenelse{\CompactFigs=0}
   {\includegraphics[width=18cm]{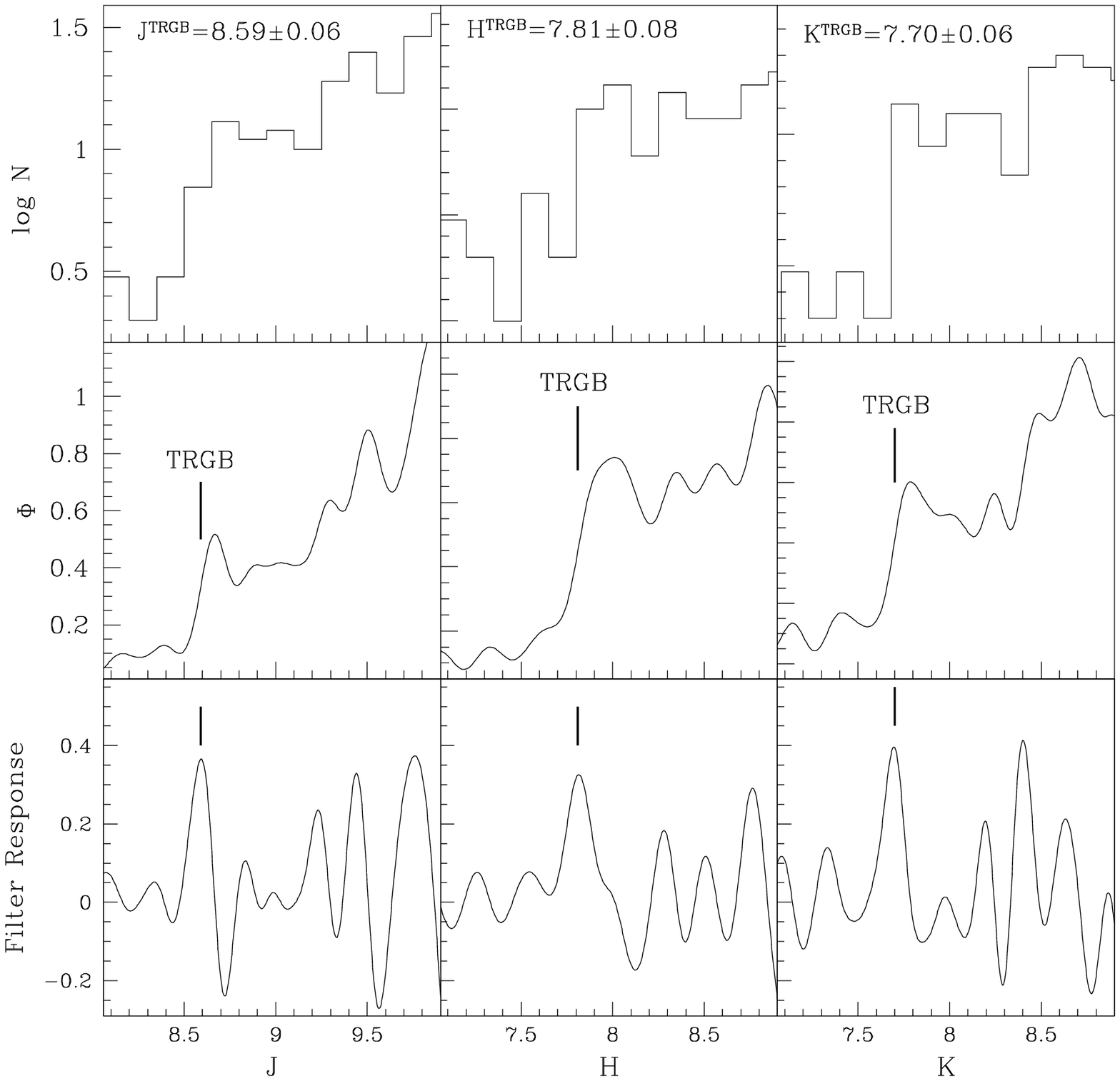}}{}
   \caption{Detection of the TRGB in J,H and K bands from the upper RGB LF 
   of $\omega$~Centauri. The upper panels report the ordinary
   log histogram, middle panels shows generalized histogram
   \cite[e.g. the histogram convolved with a gaussian with standard deviation
   equal to the photometric error at the given magnitude, see][for definitions 
   and references]{draco} and lower panels
   display the response of the Sobel's filter to the observed LF. The position
   of the TRGB is indicated by a thick vertical segment.}
    \end{figure*}
%

\subsection{Detection of the TRGB}

In the color range $-0.3<J-K<1.3$ the final catalogue
contains 72419 sources with $6.5<K<18$ taken from S03 and 47000 sources with
$6.5<K\simlt 16$, i.e. a total of $\sim 120000$ stars. 
The RGB stars brighter than $K=11$ are 2035, 940 from S03 and 1456 from the
2MASS catalogue. 

The resulting NIR Color Magnitude Diagrams (CMDs) are presented in Fig.~1. The
RGB of $\omega$~Cen is the most evident feature of these diagrams at 
$0.5\simlt J-K \simlt 1$ and ranging from $K\sim 14 $ to $K\sim 7.5$. The Blue
Horizontal branch of the cluster is also visible at $K<13$ and $J-K\simlt 0.4$.
The Asymptotic Giant Branch does not appear clearly in these diagrams because of
the unfavorable scale but it is also clearly identified in the database (see S03
for enlarged CMD).
The blue plume around $J-K\sim 0.3$ and extending to $K\sim 8$ is due to
foreground Main Sequence stars of the Galactic Disc. The level of the TRGB (as
derived below) is also reported in all CMDs of Fig.~1. Approximately 
half of the red sources ($J-K\sim 1$) brighter than the TRGB are expected to be
foreground field sources, according to the predictions of the Galactic model by
\citet{robin}. Some of them may be blended sources,
since the large majority comes from the low-spatial-resolution 2MASS catalogue.
The fact that our final estimates of the TRGB level in NIR passbands are
unaffected if we limit the analysis to the high-spatial-resolution S03 sample,
ensures that the possible contamination by blended sorces from the 2MASS
catalogue is not a serious concern in the present application.
An exhaustive analysis of these stars is clearly beyond the scope of the 
present study. The key point here is that the cut-off of the RGB population
marking the location of the Tip is clearly evident in the CMD of Fig.~1 as well
as in the Luminosity Functions shown in Fig.~2 below, hence these
``supra-TRGB'' stars don't appear to affect the actual detection
of the Tip. The main argument at the
basis of the use of RGB-LF cut-off as the marker of the TRGB level is that it
is essentially insensitive to the effects of contaminating populations of any
kind, once a well sampled RGB is considered.

The detection of the TRGB in the three NIR passbands has been performed with the
same technique used in Pap-I \citep{lfm93,mf95,mf98} and is described in Fig.~2.
The upper panels of the plot shows the log LF of the upper RGB as an ordinary
histogram while middle panel shows the LF as a generalized histogram (see Pap-I
and references therein). The sharp drop in the LF is clearly visible in both
histograms and it is neatly detected by the Sobel filter whose response is
plotted in the lower panels of Fig.~2. The position of the TRGB is marked by a
thick vertical segment. The apparent magnitude of the TRGB is $J=8.59 \pm 0.06$,
$H=7.81 \pm 0.08$, and $K=7.70 \pm 0.06$, respectively. The reported
uncertainties are the half width at half maximum of the peaks of the Sobel
filter response.

Because of the larger sensitivity to metallicity 
of the TRGB in J, H and  K with respect to I, we
were forced to search the cut-off in the LF of the blue bulk of the RGB
population, excluding the reddest stars that have metallicity $[Fe/H]\simgt
-1.4$ \citep[][and references therein]{panc,keyp,panct}. Such a selection
(illustrated in the lower panel of Fig.~1)
reduced the number of RGB stars in the upper part of the LF with respect to the
analysis of Pap-I.  We made a number of simulations using subsamples of our RGB
stars to estimate the TRGB in the same way shown in Fig.~2 and we find out that
the best estimate is recovered within $\pm 0.04$ mag in more than 95 \% of the
cases if $N_{\star}>50$, independently of the considered NIR passband. For our
best estimates (e.g. those presented in Fig.~2) $N_*=100, 69, 86$ in J,H, and K,
respectively, hence it is very unlikely that they are significantly affected 
by small-sample problems. As said, the results are unchanged also if we limit 
the analysis to the S03 sample. 

Adopting the distance modulus and reddening described in the previous section,
and assuming the reddening laws by \citet{rieke} we obtain $\mj = -5.20 \pm
0.16$, $\mh = -5.94 \pm 0.18$ and $\mk = -6.04 \pm 0.16$, where the error bars
account for all the involved sources of uncertainty\footnote{In the following we
will always adopt the reddening laws by \citet{rieke} for J,H and K bands and \citet{dean}
for the Cousins' I passband.}.

%
   \begin{figure*}
   \centering
   \ifthenelse{\CompactFigs=0}
   {\includegraphics[width=18cm]{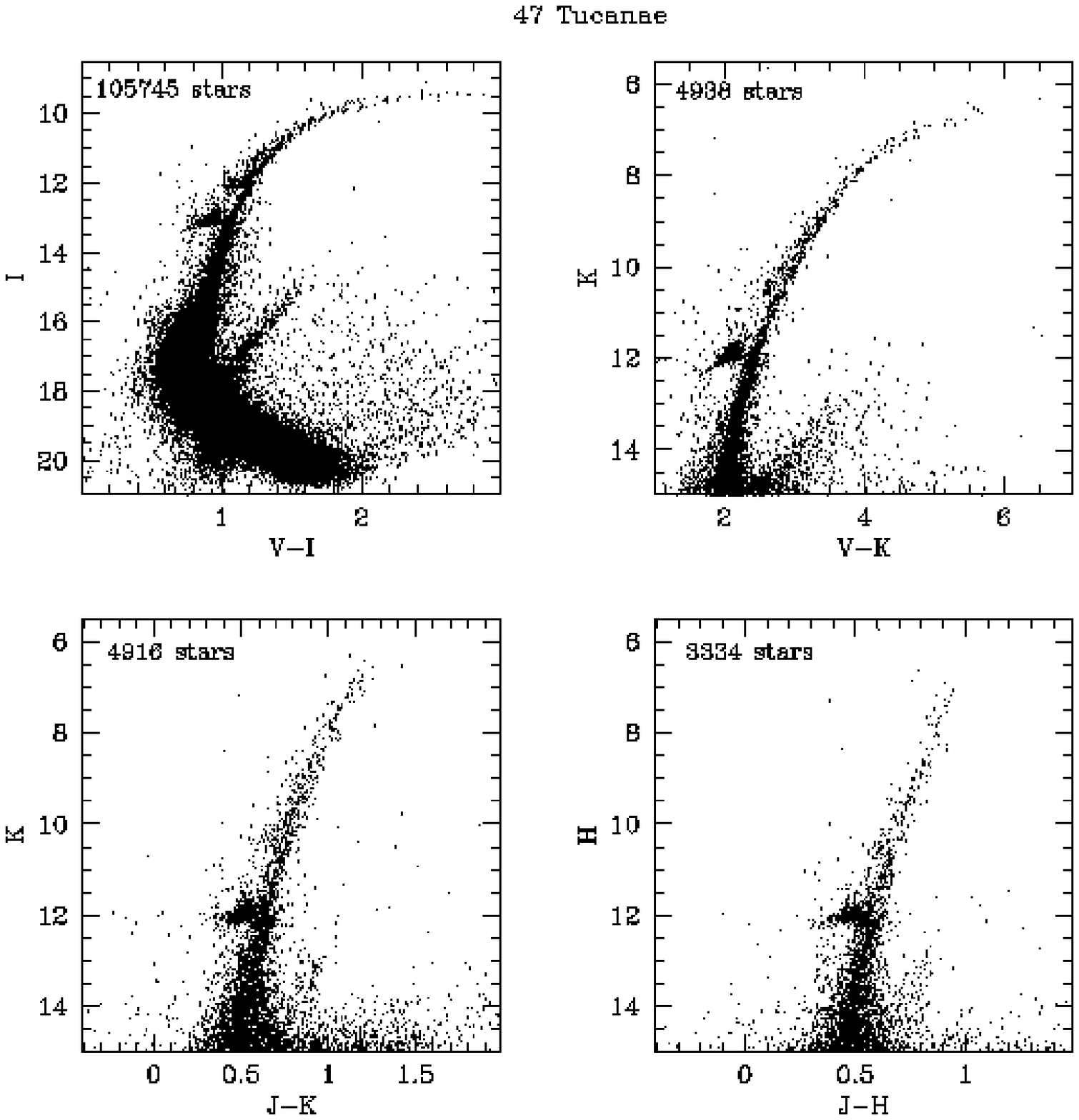}}
   \caption{Optical and NIR CMDs of 47~Tucanae obtained from
   our photometric sample implemented with the 2Mass database and the J,K
   photometry by \citet{paolo}. The feature at $I>14$ and $K>12$, mostly to the
   red of the cluster CMD is the RGB of the Small Magellanic Cloud that lies in
   the background.}
    \end{figure*}
%

\section{The TRGB of 47~Tuc in I,J,H and K}

We retrieved from the ESO archive seven 60s images for each filter (B,V,I)
centered on the center of 47 Tucanae. The images were taken with the ESO-MPI-WFI
camera mounted at the 2.2~m telescope at La Silla (Chile) as part of the 
Pre-Flames ESO Imaging Survey 
(EIS)\footnote{http://www.eso.org/science/eis/surveys/strategy\_Pre-Flames.html}.
Since the upper RGB stars of the cluster were saturated in the EIS images, we
complemented the observational material with one proprietary 20 s exposure for 
each filter taken with the same instrument. The seeing of the considered images
ranges from $0.8\arcsec$ to $1.5 \arcsec$ with a median of $\sim 1.1 \arcsec$
The images were corrected for bias
and flat-field with standard IRAF procedures and the photometry of each of them
was independently obtained using DoPhot \citep{dophot}. For each passband
the catalogues from the seven 60 s exposures were cross-correlated, reported to
the same (instrumental) photometric system and the stars
in common were averaged with the additional requirement that a star is included
in the final catalogue only if it is successfully measured at least in 3 images
per filter. The averaged B,V,I catalogues were finally combined and merged with
the catalogue obtained from the short exposures, after reporting the latter to
the same photometric system. All the stars in the total catalogue with 
$I\simlt12$ are from the 60 s exposures while those brighter than this limit are
from the 20 s exposure. The long/short exposure threshold is approximately
located at the base of the AGB clump. The absolute calibration was obtained by
direct comparison with the large set of secondary standard stars provided
by \citet{stet} that are present in our field. More details about the dataset
will be provided in a future paper devoted to a more thorough study of the
cluster, while here we are interested only in the TRGB. The final optical
catalogue contains 105745 stars in the range $0.0<V-I<3.0$ and $8.5\le I <21$.

The NIR dataset was assembled in a similar way to that for $\omega$~Cen
described above. J,K photometry for the inner $4\times 4$ arcmin$^2$ has been
taken from \citet[][hereafter M95]{paolo} and J,H,K photometry of a circle
centered on the cluster of radius $r=33 \arcmin$ was retrieved from the PSC of
2MASS with the same requirements and transformations adopted for $\omega$~Cen. 
There are very few stars in
the inner $\simeq 6\arcmin$ around the center in the 2MASS catalogue, hence
there is a nearly annular region  
between $r\sim 4\arcmin$ and $r\sim 6\arcmin$ that is not sampled by our
total NIR catalogue. Therefore, in the case of 47 Tuc, the statistics of RGB
stars in the NIR could be improved by observations that cover also this
area. In our final NIR catalogue there are more than 4000 stars with $K<15$. The
RGB stars with $K<11$ are 445, 240 of which are from M95 and 205 from the 2MASS
catalogue. Finally we identified all the Long Period Variables (LPV) from the
lists by M95 and \citet{clement} that are present in our sample (namely
V1, V2, V3, V4, V6, V7, V8, V11, V13, V19, V20, V21, V22, V26, V27, A19).

To illustrate the properties of our sample we plot in Fig.~3 some examples of
optical, optical-infrared and infrared CMDs. The RGB is well defined in every
plane and it is clearly separated from the red HB and from the AGB clump and
sequence. The red plume that crosses the cluster MS in the (I, V-I) diagram is
the RGB of the Small Magellanic Cloud that lies in the background of our field.

%
   \begin{figure}
   \centering
   \ifthenelse{\CompactFigs=0}
   {\includegraphics[width=9cm]{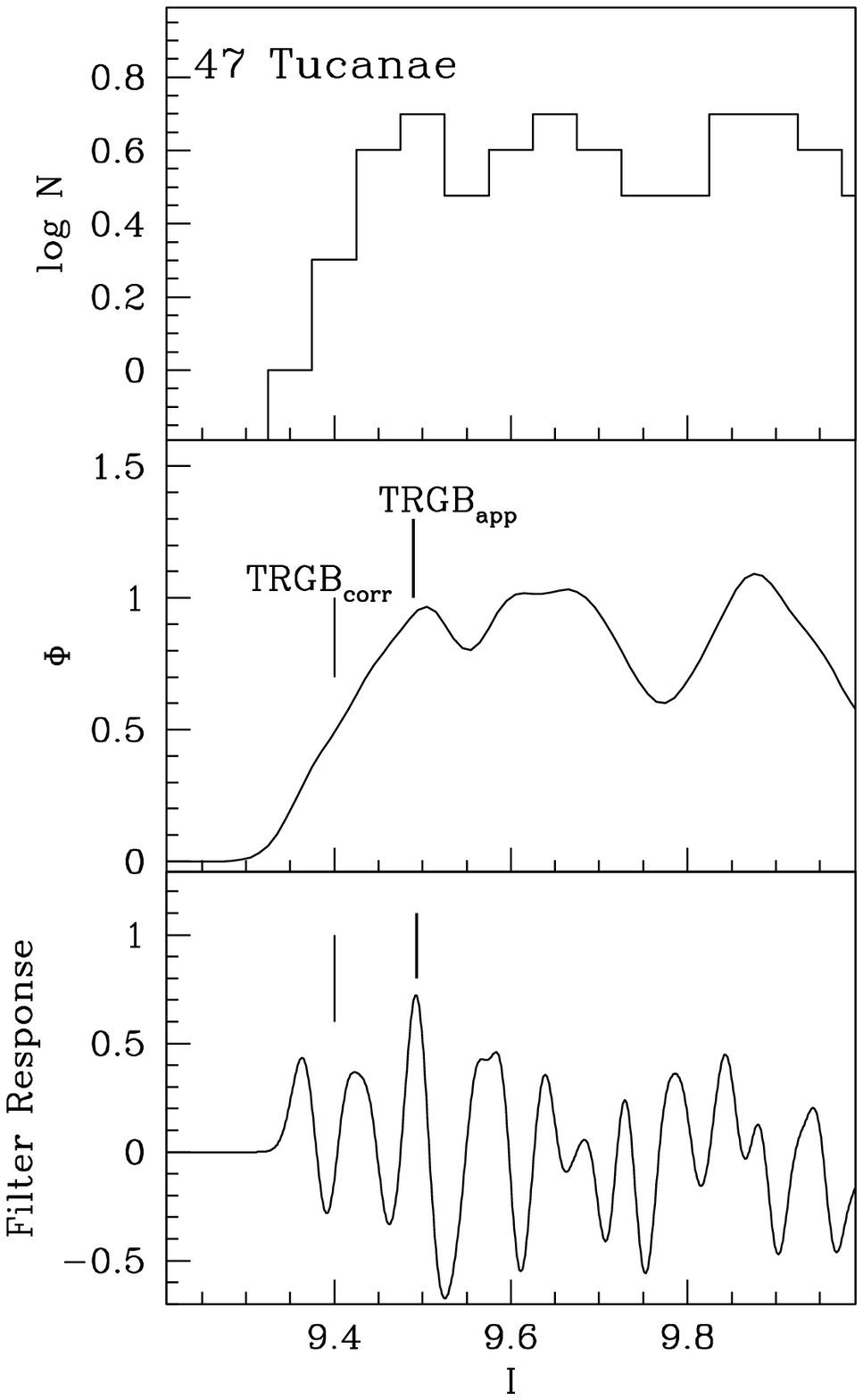}}{}
   \caption{Detection of the TRGB in I for 47~Tucanae, with the same technique
   displayed in Fig.~2. The apparent (TRGB$_{app}$) and corrected 
   (TRGB$_{corr}$) positions of the TRGB are also shown.}
    \end{figure}
%

\subsection{Detection of the TRGB}

Fig.~4 illustrates the detection of the TRGB in the I passband, obtained with
the same technique adopted in Pap-I and in Fig.~2. 
We find $I_{TRGB}=9.49 \pm 0.1$.
Since $N_{\star}=80$ we have to correct our estimate with the relation provided
by \citet{draco}. In the present case the correction is 
$\delta I_{TRGB}=-0.09 \pm 0.08$,  hence our final estimate is  
$I_{TRGB}=9.40 \pm 0.13$. The corrected position of the TRGB is also reported in
Fig.~4.

%
   \begin{figure*}
   \centering
   \ifthenelse{\CompactFigs=0}
   {\includegraphics[width=18cm]{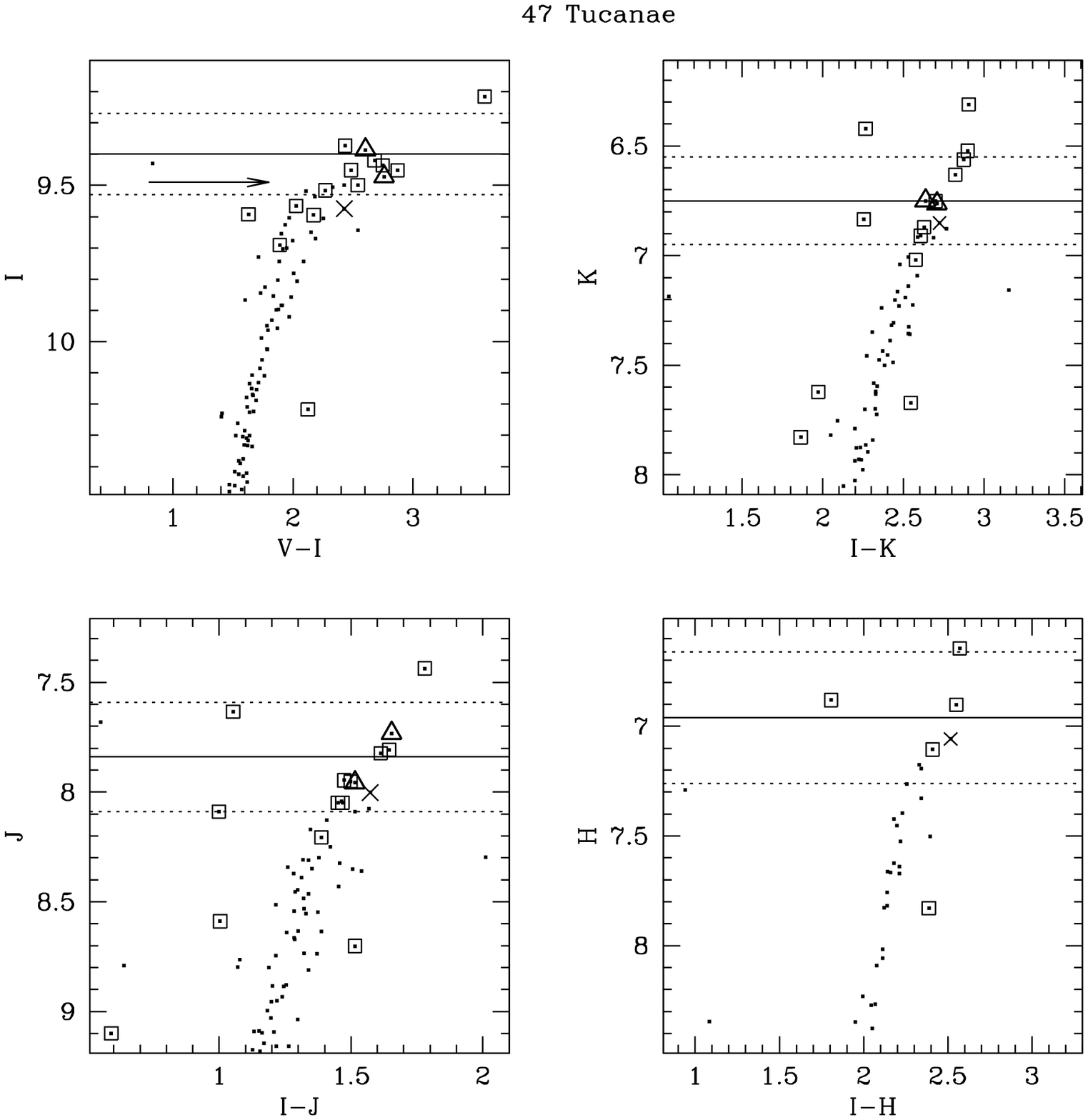}}{}
   \caption{The upper RGB of 47~Tucanae is shown in CMDS obtained from various
   combinations of the I,K,J, and H photometry. 
   The arrow in the (I,V-I) CMD marks the
   position at which the TRGB is detected by the Sobel's filter (Fig.~4) while
   the continuous line marks the {\em true} position of the TRGB once the
   correction for the poor sample is applied. The dotted lines enclose the
   $\pm 1\sigma$ uncertainty range around this level. Open squares are long
   period variables. The open triangles are the brightest non-variable stars
   that have valid photometry in I,J and K and the cross is the  
   brightest non-variable star that has I and H photometry.}
    \end{figure*}
%
%

The NIR database for 47~Tuc is significantly smaller than the optical one and
the number of RGB stars sampled is insufficient for a safe application of the
technique adopted above. Hence we are forced to use the brightest non-variable
RGB star as a tracer of the TRGB. However, in doing this, we take advantage of
the robust detection we obtained in I by imposing self-consistency
among the estimates in the various passbands. In practice, we require that the
stars marking the position of the TRGB shall be the same in any of the 
considered filters.

The followed procedure is illustrated in Fig.~5 in which the CMDs of the upper RGB
are plotted in the various optical/IR planes. 
The are two
non-variable stars which lie very close to the position of
the TRGB. We use these two stars as tracers of the TRGB position in K and J,
taking their average magnitude as the best estimate for the TRGB and adding in
quadrature the standard deviation to the uncertainty in $I_{TRGB}$.
Note that in the K passband the two stars have nearly 
the same magnitude. Note also that the selected stars are {\em the
brightest non-variable stars} in any of the considered passbands. Hence our
choice of the ``TRGB markers'' is fully justified independently of the
operational definition actually adopted, e.g. the RGB-LF cut-off or the
brightest non-variable star.

Unfortunately these stars lack H photometry. The brightest
non-variable star with H photometry is marked by a cross in the CMDs of Fig.~5.
It can be appreciated that this star is 0.1 mag fainter than the TRGB 
level in K. It is reasonable to assume that a similar underestimate of the TRGB
level should occur also in the H band, therefore, as our best estimate of
$H_{TRGB}$ we take the H magnitude of the considered star shifted
by -0.1 mag.

It is obvious that the estimates of the observed TRGB level of 47~Tuc
described above suffer from larger uncertainties with respect to
the case of $\omega$ Cen. We take into account this
fact by conservatively combining all the sources of uncertainty in the final
calibrations, including the uncertainty associated with $I_{TRGB}$ which drove our
detections in the NIR bands. Our final estimates are: $J_{TRGB}=7.84 \pm 0.25$,
 $H_{TRGB}=6.96 \pm 0.30$, and $K_{TRGB}=6.75 \pm 0.20$. 
 
\subsection{Comparison with previous analysis}
 
On the other hand it is interesting to consider the impact of the adoption of
our large database on the estimate of the TRGB level from a purely observational
point of view. The brightest non-variable star of 47~Tuc in the original DA90
sample has I$=9.65$, while in the final sample adopted by these authors
(implemented with photometry from other sources) the brightest non-variable  
star has I$=9.50$, that is 0.25 and 0.1 mag fainter, respectively, than our 
{\em observed} value    
(I$=9.40$), before any correction for under-sampling ($\delta I$). If the
corrected estimate is considered (I$=9.33$), it has to be concluded that 
previous
estimate of $I_{TRGB}$ for 47~Tuc missed the true TRGB level by $\sim 0.2$ mag
because of the severe under-sampling of the upper RGB population (see Pap-I and
references therein). The effect is not so large for the NIR passband, since
the dataset of \citet{paolo} contains many more RGB stars with respect to the
optical sample used by DA90 and the bulk of the evolved stars in our catalogue
are taken from \citet{paolo}. Our estimate of $K_{TRGB}$ coincides with that by
\citet{paolo}, while our $J_{TRGB}$ is 0.1 mag brighter with respect to their
estimate. No previous estimate of $H_{TRGB}$ is available in the literature.
Hence, despite the considerable uncertainties that still affect the 
estimates of the TRGB level of 47~Tuc presented in this paper, they represent 
a significant progress with respect to those previously available. 

The above example shows in practice that the derivation of $I_{TRGB}$ from 
small samples may provide serious underestimation of the TRGB luminosity. 
Hence, it may appear
surprising that the $M_{TRGB}^I$ calibration by DA90 - whose original estimates
of the TRGB level are based on small samples - is in good agreement with the
calibration of Pap-I, whose zero-point relies on the huge sample of
$\omega$~Cen stars by \citet{panc}. This fact may be ascribed to the
different distance scales adopted in the two cases. For example, the distance
modulus of 47~Tuc adopted by DA90 is $(m-M)_0=13.39$, $0.08$ mag larger that
what assumed here (see below) and $0.11$ mag larger than what reported by F99.
This difference in the distance scale may partially compensate the effect of
the observational under-sampling, reconciling the two calibrations to similar
results, notwithstanding the difference in the {\em observed} TRGB magnitude.

   \begin{figure}
   \centering
   \ifthenelse{\CompactFigs=0}
   {\includegraphics[width=9cm]{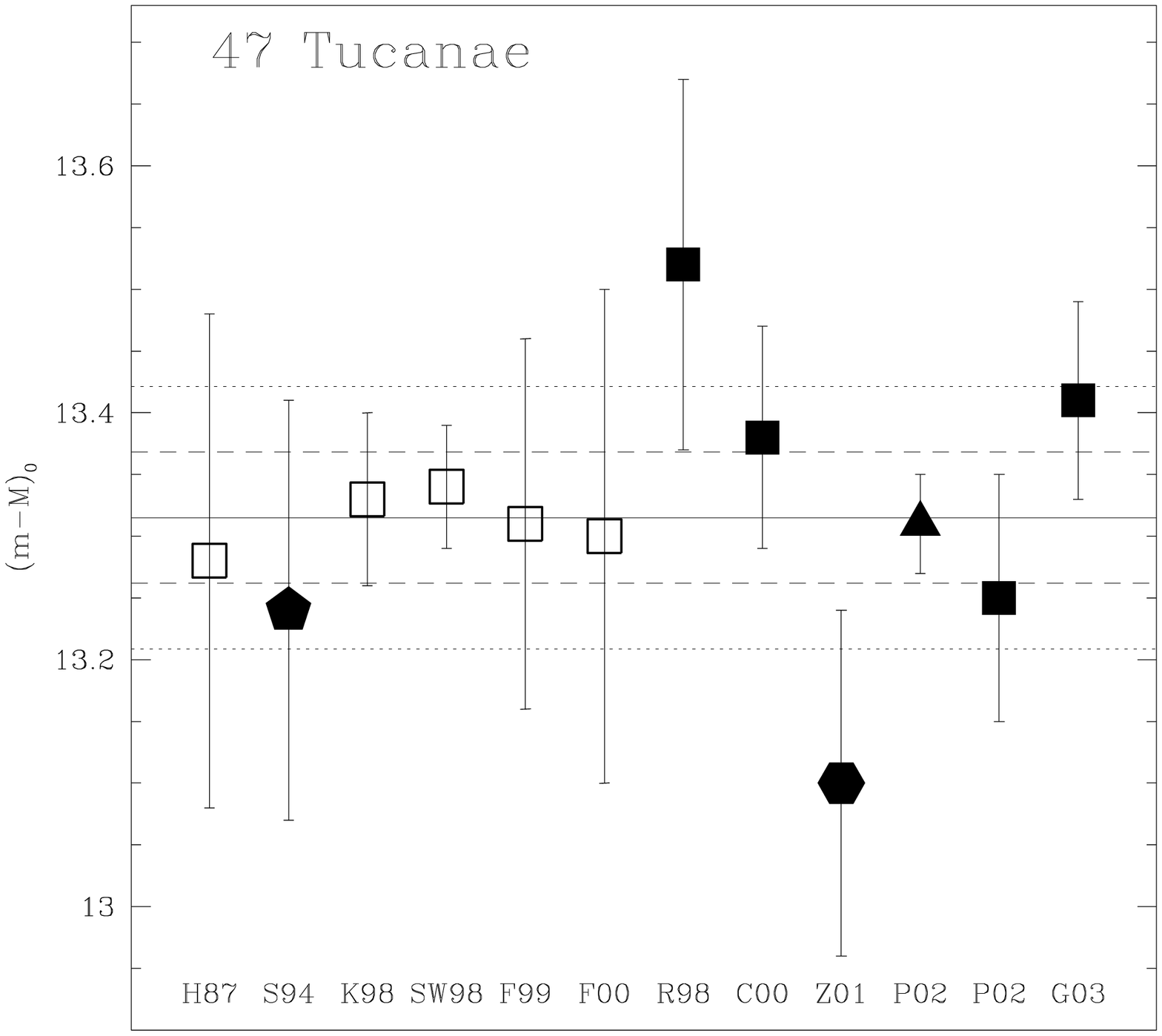}}{}
   \caption{Summary of recent estimates of the true distance modulus of
   $47~Tucanae$. The open symbols refers to estimates relying on the RR Lyrae/HB
   distance scale while the filled symbols are from independent methods. In
   particular: filled squares = main sequence fitting; filled triangles = red
   clump (Hipparcos-calibrated); filled pentagons = Baade-Wesselink; filled
   hexagons = white dwarf sequence fitting. The continuous line is the mean of
   the estimates that are independent of the RR Lyrae distance scale, the dashed
   and dotted lines enclose the $\pm 1\sigma$ and $\pm 2\sigma$ ranges about
   the mean, respectively.}
    \end{figure}
%

\subsection{Distance modulus}

Though 47~Tuc is a nearby and very well studied cluster, 
there is significant scatter in the distance modulus estimates obtained by
various authors with different methods 
\cite[see][for discussions]{perci,manu,g03}. 
The situation regarding the most recent
determinations is clearly illustrated in Fig.~6, where the open symbols marks
estimates based on (various) RR Lyrae/HB distance scales while filled symbols
represent estimates that are independent of such scale. It is clear that there
is little basis to prefer any given RR Lyrae/HB-independent distance estimate to
the others. As a compromise solution, we adopt the
average of the HB-independent estimates as our final $(m-M)_0$ value and the
standard deviation as its uncertainty, reported as continuous and dashed lines
in Fig.~6: $(m-M)_0=13.31\pm0.14$. We just note here that the adopted distance
modulus is in good agreement with the estimates by F99 and F00 whose distance
scale is fully compatible with our TRGB scale, as shown in Pap-I. For
consistency with F99 we also adopt E(B-V)$=0.04$, also in agreement with the
most recent determination by \citet{g03} from the B-V index (see their Tab.~3).

\section{Calibrations}

Following F99 and F00 we provide the calibration of the TRGB level in I,J,H and
K bands as a function of the {\em global metallicity} [M/H]. 
This parameter includes
the contribution of the $\alpha$-elements according to the formula provided by
\citet{scs93}. The observed properties of a given stellar population
(at fixed age and helium content) do not depend simply on
the iron abundance but also on the abundance of the $\alpha$-elements 
\cite[see][for discussion and references]{scs93,f99,trag}, 
and some variations of the [$\alpha$/Fe] abundance ratio are observed among globulars
\cite[see, for example][]{carney,brown} and galaxies 
\cite[see][and references therein]{shet,bonicaf,trag}.
For these reasons, [M/H] provides a 
more suitable indicator of the global metal content 
to relate with the observed properties 
of stars and stellar populations. 
We assume [$\alpha$/Fe]=$+0.35$ for $\omega$~Cen, according to the recent 
review by \citet{panct}, and [$\alpha$/Fe]=$+0.2$ for 47~Tuc, following 
\citet{carney} and the more recent results by \citet{c03}. The final values of 
the global metallicity (computed with Eq.~1 in F99) are [M/H]=$-1.45\pm 0.1$ 
and [M/H]=$-0.58\pm 0.1$, for $\omega$~Cen and 47~Tuc, respectively.

The final calibrations are reported in Fig.~7. All the plots are in the same
scale to better appreciate the different dependence on metallicity and the
different degree of uncertainty of the calibrations in the various passbands.
The two observational points
derived here are compared with the predictions of the
theoretical models by 
\citet[][kindly provided in electronic 
form by Santino Cassisi]{sc98}. Panel (a) of Fig.~7 shows the calibration
in the I band. The quoted relation is the same of Pap-I, with a 
zero-point shift of $+0.03$ mag that we applied to achieve the best
match to the observed points (thick continuous line). From the inspection of
panel (b) it can be appreciated that the empirical relation for $M_K^{TRGB}$
obtained by F00 provides a good fit to the
observed points\footnote{As discussed in Pap-I, the RGB samples adopted by
F00 to detect the TRGB in the K passband are much richer than those used
by DA90, for instance (see Sect.~4.2). Hence their estimates of the TRGB level
are quite reliable, in general, and it is not surprising that their empirical
relation is in good agreement with the accurate zero-points derived here.
However, it has to be noted that the F00 relation is 0.05 mag fainter, in K,
with respect to our $\omega$ Cen zero-point and the overall uncertainty in the
single TRGB detections are significantly larger than what we obtained here
for $\omega$ Cen and 47 Tuc. Finally, as also discussed in Pap-I, in the attempt
to provide calibrating relations for extragalactic applications of the TRGB
technique it is relevant, in our view, that the TRGB detection for the pillars
of the calibration is based on the same operational definition of the observable
that is used when dealing with distant galaxies, e.g. the cut-off of the RGB LF.
This operational definition can be applied only to large RGB samples as those
described in Sect.~2.2.}.
Hence we adopt it as our final calibration in the K band.
There are no available empirical calibrations of the TRGB in the J and H
bands. 
However, since the considered models are in excellent agreement with our
pillar references, we adopt a linear fit to the theoretical points as 
our final calibrations in this passbands (Fig.~7, panels (c) and (d)).
This choice guarantees a full consistency
with the observational constraints as well as a sound basis to extrapolate the
calibration outside of the metallicity range enclosed by the two considered data
points.

The disagreement between our empirical 
calibration and the theoretical models by \citet{sc98} in the I band, 
already described and discussed in Pap-I, is quite evident also in
Fig.~7(a). It is interesting to note that, on the other hand, models predictions
are in excellent agreement with the observed points in all three NIR passband
considered. This occurrence demonstrates that the difference found in the I band
is not due to a problem in our distance scale, since the reddening and distance
assumptions are the same for all the passbands, nor to an underestimate of the
observed luminosity of the TRGB in I due to under-sampling, since the available
optical sample contains many more RGB stars with respect to the NIR one. At the
present stage, we can only suggest that the most likely origin of the
disagreement may reside in unsatisfactory color transformation of theoretical 
models in the I band. Note, however, that there are also intrinsic ($M_{bol}$) 
differences in the TRGB level prediction between the various theoretical models
\cite[see][for references and discussion]{sal}.

   \begin{figure*}
   \centering
   \ifthenelse{\CompactFigs=0}
   {\includegraphics[width=18cm]{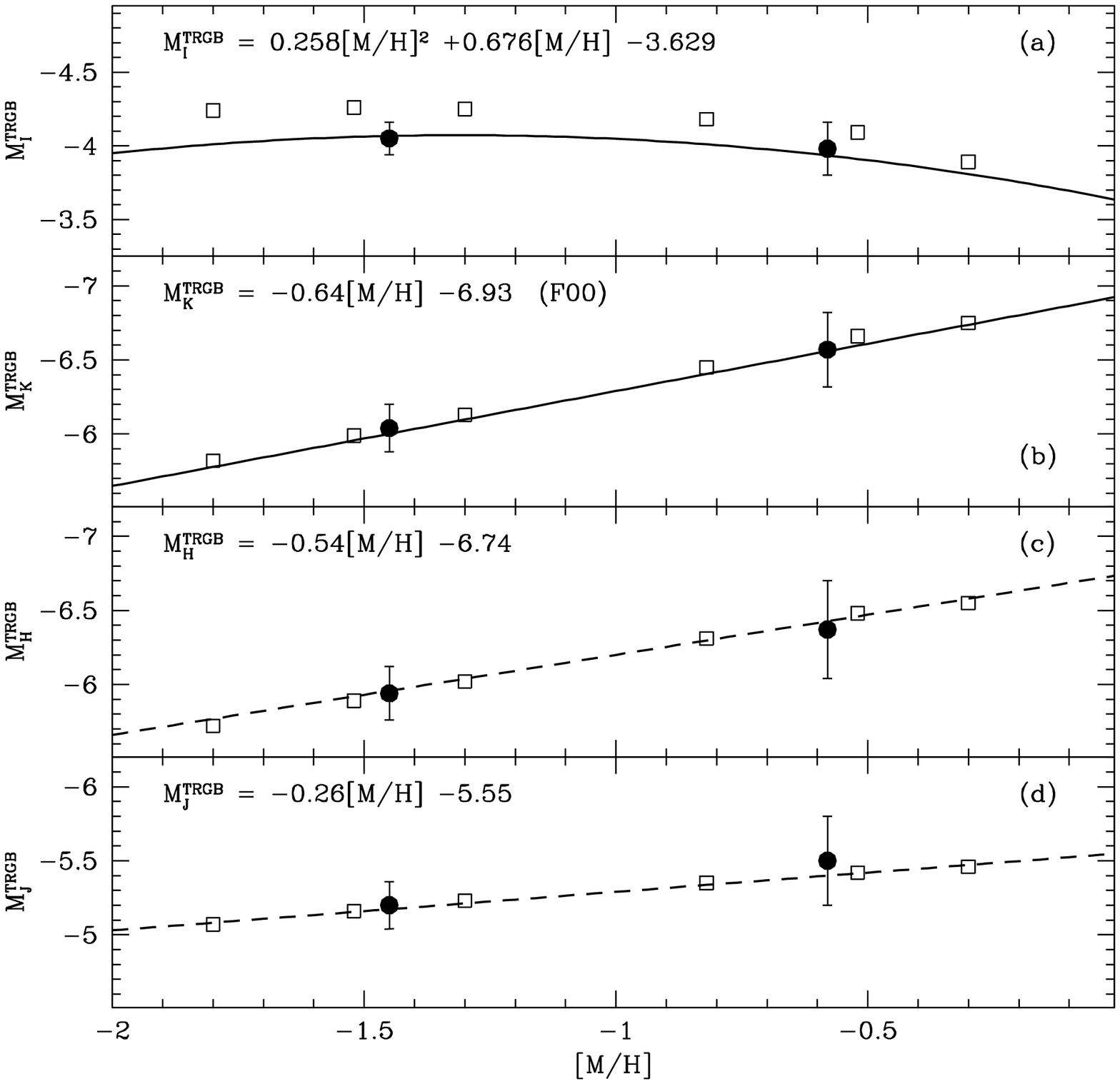}}{}
   \caption{Calibration of the absolute I,K,H,J magnitude of the TRGB as a
   function of the global metallicity. 
   The filled circles are the observational
   points, the open squares are the theoretical models by \citet{sc98}. 
   {\it Panel~(a)}: the continuous line is the empirical calibration derived in Pap-I 
   shifted by +0.03~mag (see Sect.~5).
   {\it Panel~(b)}: the continuous line is the empirical calibration by F00.
   {\it Panels~(c,d)}: the dashed lines are the linear fit to the model points.
   }
    \end{figure*}
%
%

\subsection{The TRGB distance scale: the case of LMC}

The distance to the Large Magellanic Cloud (LMC) is the fundamental keystone of
the extragalactic distance ladder and it has been the subject of considerable
debate in the last decades. 
However, in recent years most of the results appear to converge toward a
``standard value'' $(m-M)_0\simeq 18.50$ \citep{alves2,gisella}. Hence, it is
interesting to check if our independent TRGB distance scale provides a distance
modulus of the LMC that is consistent with the standard value or not. 

To do that, we take the observed apparent TRGB level in I,J and K provided by the
DENIS survey \cite[][hereafter Cal00]{denis} and we adopt [M/H]=$-0.7$ for the
LMC as done by the same authors. Given these assumptions we obtain an estimate
of the true distance modulus and reddening from each couple of calibrating
relations in the available passbands, i.e. I and J, I and K, J and K by finding
the value of the reddening for which (exactly) the same $(m-M)_0$ is obtained 
from the two considered relations. The average and standard deviations of the 
three estimates obtained in this way are $<(m-M)_0>=18.46 \pm 0.06$ and 
$<E(B-V)>=0.07\pm 0.06$. As expected, given the considerable uncertainties
affecting the NIR relations presented in Fig.~7, 
the constraint on the reddening are quite poor but the (self-consistent)
obtained distance modulus is in line with the standard value. 

It is interesting to note that if we assume $E(B-V)=0.15$, as done by Cal00,
the distance modulus we obtain from $I_{TRGB}$ is considerably smaller than
what obtained in the other passbands (18.28 with respect to 18.43 and 18.42, in
I, J, and K respectively). A full self consistency among the various passbands 
is achieved only if a smaller value of the reddening is adopted. However there
are serious arguments favoring the adoption of a lower extinction 
\cite[see][]{zarebv,gisella,dutra}.
In particular, 
\citet{zarebv} has shown how critical may be the reddening issue in the estimate
of the LMC distance modulus. According to this author, not only the amount of 
extinction varies significantly with the location within the galaxy but also as
a function of the temperature of the considered stars. In particular, he
demonstrated that the cool stars (as RGB stars) are less extincted than hot
stars 
\citep[essentially young Main Sequence stars, see][for further details]{zarebv}.
According to \citet{obmap} and \citet{zarebv} the reddening adopted by Cal00 is
probably more appropriate for OB stars than for the cool RGB we are considering.

\citet{sakai} provide a robust estimate of the de-reddened apparent TRGB
magnitude using a very large sample of LMC RGB stars. The observed I magnitude
of each single star is corrected for reddening using the accurate extinction map
provided by \citet{obmap} and \citet{zarebv}, and taking into account the 
temperature dependence of the extinction according to the prescriptions 
of \citet{zarebv}.
Adopting their result $I^0_{TRGB}=14.54 \pm 0.04$ we obtain 
$(m-M)_0=18.54 \pm 0.04$ (statistic) $\pm 0.12$ (systematic).

We conclude that our TRGB distance scale, whose zero-point is based on the
distance to $\omega$~Cen derived from the detached eclipsing binary OGLE-17,
provides a distance modulus in full agreement with the standard value 
$(m-M)_0\simeq 18.50$. The same conclusion has been reached by \citet{alves2}
using an independent dataset and the original I calibration we provided in
Pap-I, which is essentially the same shown in Fig.~7a, here. The same
result is obtained independently of the adopted passband, once the correct
reddening value is adopted, showing that the
different calibration shown in Fig.~7 are fully self-consistent. 

\section{Summary and Conclusions}

We have refined the zero-point of the calibration of the TRGB as a standard
candle provided by \citet[][Pap-I]{tip} adopting a more accurate estimate of the
reddening to $\omega$ Centauri, the fundamental pillar of our distance scale.
The calibration of Pap-I has been accordingly revised and extended to NIR
(J,H,K) passbands using a new large photometric dataset of $\omega$ Cen giants
taken from S03 and 2MASS.

A large optical and NIR photometric database has been assembled for the metal
rich globular cluster 47~Tucanae. From this database we obtained the best
estimate of the TRGB presently available for this cluster in I, J, H and K,
providing a second observational point for our TRGB calibrations, at
$[M/H]\simeq -0.6$.

With this new observational material we provide new robust calibrations of the
TRGB magnitude in I, J, H and K bands as a function of the global metallicity. 
The obtained NIR calibrations are in excellent
agreement with the theoretical predictions by \citet{sc98}, while the marginal
discrepancy in the I band already noted in Pap-I persists and it is probably due
to the color transformation applied to theoretical models.
It has also been shown that our calibrations self-consistently provide a
distance modulus of the Large Magellanic Cloud in good agreement with the
``standard value'' \cite[see][]{alves2,walk,gisella}.

With the present contribution, we think we have provided the ``state of the art''
instrument of stellar astrophysics needed to afford the study of extragalactic
distances with the Population II distance scale \citep{walk,alves2}, e.g. a
fundamental complement and consistency test for Cepheid and SNIa distance
scales.

\begin{acknowledgements}
We acknowledge the  financial support to this research by
the Italian {Ministero  dell'Universit\'a e della Ricerca Scientifica}
(MURST). We are grateful to S. Cassisi for providing models predictions
in electronic form and for many insightful discussion.

Part of the data analysis has been performed with software developed by P.
Montegriffo at the Osservatorio Astronomico di Bologna (INAF).  
This research has made use of ESO Archive data taken with the WFI 
at the MPI/ESO 2.2m Telescope at the 
La Silla Observatory.
This publication makes use of data products from the Two Micron All 
Sky Survey, which is a joint project of the University of Massachusetts and 
the Infrared Processing and Analysis Center/California Institute of 
Technology, funded by the National Aeronautics and Space Administration and 
the National Science Foundation.
This research has made use of NASA's Astrophysics Data System Abstract
Service. 
\end{acknowledgements}

{}


\begin{thebibliography}{}
\bibitem[Alves (2003a)]{alves} Alves, D., 2003a, in Stellar Candles for the 
         Extragalactic Distance Scale, D. Alloin and W. Gieren Eds., Springer,
	 Lecture Notes in Physics, in press (astro-ph/0303035)
\bibitem[Alves (2003b)]{alves2} Alves, D., 2003b, in Extragalactic Binaries,
         IAU XXV General Assembly, A. Gimenez and I. Ribas Eds., New Astronomy 
	 Review, in press (astro-ph/0310673)
\bibitem[Buzzoni (1989)]{buz} Buzzoni A., 1989, ApJ Suppl., 71, 817	 
\bibitem[Buzzoni (2002)]{buz2} Buzzoni A., 2002, AJ, 123, 1188	 
\bibitem[Barker et al. (2004)]{barker} Barker M.K., Sarajedini A., Harris J.,
         2004, ApJ Suppl., in press (astro-ph/0401387) 	 
\bibitem[Bellazzini et al. (2002)]{draco} Bellazzini, M., Ferraro, F.R.,
         Origlia, L., Pancino, E., Monaco, L., Oliva, E., 2002, AJ, 124, 3222
\bibitem[Bellazzini, Ferraro \& Pancino(2001)]{tip} Bellazzini, M., Ferraro,
         F.R., Pancino, E., 2001, ApJ, 556, 635	
\bibitem[Bonifacio \& Caffau (2003)]{bonicaf} Bonifacio P., Caffau E., 2003,
         A\&A, 399, 1183	   
\bibitem[Brown, Wallerstein \& Zucker (1997)]{brown}Brown J.A., Wallerstein G., 
         Zucker D., 1997, AJ, 114, 180	 	 
\bibitem[Cacciari(1999)]{carla} Cacciari, C., 1999, in Harmonizing Cosmic 
         Distance Scales in a Post-HIPPARCOS Era, D. Egret and A. Heck Eds.,
	 ASP, S. Francisco, ASP Conf. Ser., vol. 167, p. 140
\bibitem[Caputo, Degl'Innocenti \& Marconi(2002)]{caputo} Caputo, F.,
         Degl'Innocenti, S., Marconi, M., 2002, in $\omega$ Centauri: A Unique
	 Window into Astrophysics, F. van Leeuwen, J. Hughes and G. Piotto Eds.,
	 S. Francisco, ASP, ASP Conf. Series, 265, 185	
\bibitem[Carney (1996)]{carney} Carney B.W., 1996, PASP, 108, 900	  
\bibitem[Carpenter (2001)]{carp} Carpenter, J.M. 2001, AJ, 121, 2851
\bibitem[Carretta \& Gratton(1997)]{gc97} Carretta, E., Gratton, R., 1997,
         A\&AS, 121, 95
\bibitem[Carretta et al.(2000)]{c00} Carretta, E., Gratton, R. G., Clementini, G., 
         Fusi Pecci, F. 2000, ApJ, 533, 215 (C00)
\bibitem[Carretta et al.(2003)]{c03} Carretta E., Gratton R.G., Bragaglia A.,
         Bonifacio P., Pasquini L., 2003, A\&A, in press (astro-ph/0311347)	 
\bibitem[Cioni et al.(2000)]{denis} Cioni M.-R.L., van der Marel R.P., Loup C.,
         Habing H.J., 2000, A\&A, 359, 601	 
\bibitem[Clement (1997)]{clement} Clement, C., 1997, AAS Newsl., 84, 15	
\bibitem[Clementini et al.(2003)]{gisella}Clementini G., Gratton R., Bragaglia
         A., Carretta E., Di Fabrizio L., Maio M., 2003, AJ, 125, 1309 	 	 
\bibitem[Cutri et al.(2003)]{cutri} Cutri et al., 2003, Explanatory Supplement 
         to the 2MASS All Sky Data Release\footnote{ 
{\tt http://www.ipac.caltech.edu/2mass/releases/allsky/doc/explsup.html}}
\bibitem[Da Costa \& Armandroff(1990)]{da90} Da Costa, G.S., 
         Armandroff, T.A., 1990, AJ, 100, 162 (DA90)
\bibitem[Dean, Warren \& Cousins(1978)]{dean} Dean, J.F., Warren, P.R., 
         Cousins, A.W.J., 1978, MNRAS, 183, 569
\bibitem[Dutra et al. (2001)]{dutra} Dutra C.M., Bica E., Clari\'a J.J., Piatti
         A.E., Ahumada A.V., 2001, A\&A, 371, 895	 
\bibitem[Elias (1983)]{elias} Elias, J.H., Frogel, J.A., Hyland, A.R., \&
         Jones, T.J.  1983, AJ, 88, 1027
\bibitem[Ferraro et al.(1999)]{f99} Ferraro, F.R., Messineo, M., Fusi Pecci,
         F., De Palo, M.A., Straniero, O., Chieffi, A., Limongi, M., 1999, 
	 \aj, 118, 1738 (F99)
\bibitem[Ferraro et al.(2000)]{f00} Ferraro, F.R., Montegriffo, P., Origlia,
         L., Fusi Pecci, F., 2000, \aj, 119, 1282 (F00)
\bibitem[Ferraro, Pancino \& Bellazzini(2002)]{keyp} Ferraro, F.R., Pancino, E.,
         Bellazzini, M., 2002, in $\omega$ Centauri: A Unique
	 Window into Astrophysics, F. van Leeuwen, J. Hughes and G. Piotto Eds.,
	 S. Francisco, ASP, ASP Conf. Series, 265, 407 	 
\bibitem[Gratton et al.(2003)]{g03} Gratton, R.G., Bragaglia, A., Carretta, E.,
         Clementini, G., Desidera, S., Grundahl, F., Lucatello, S., 2003, A\&A, in
	 press (astro-ph/0307016, G03)
\bibitem[Harris(1996)]{harris} Harris, W.E., 1996, AJ, 112, 1487
\bibitem[Harris, Zaritsky \& Thompson(1997)]{obmap} Harris J., Zaritsky D.,
         Thompson I., 1997, AJ, 114, 1002	 	 
\bibitem[Hesser et al.(1987)]{h87} Hesser, J. E., Harris, W. E., VandenBerg, D. A., 
         Allwright, J. W. B., Shott, P., Stetson, P. B. 1987, PASP, 99,	739 (H87)
\bibitem[Hilker \& Richtler (2000)]{hilker} Hilker M., Richtler T., 2000,
         A\&A, 362, 895
\bibitem[Hughes \& Wallerstein (2000)]{hughes} Hughes J., Wallerstein G., 2000,
         AJ, 119, 1225	 
\bibitem[Kaluzny et al.(1998)]{k98} Kaluzny, J., Wysocka, A., Stanek, K. Z., 
         Krzeminski, W., 1998, Acta Astron., 48, 439 (K98)	  
\bibitem[Kov\`acs (2002)]{kovacs} Kov\`acs, G. 2002, in $\omega$ Centauri: A Unique
	 Window into Astrophysics, F. van Leeuwen, J. Hughes and G. Piotto Eds.,
	 S. Francisco, ASP, ASP Conf. Series, 265, 	 
\bibitem[Lee, Demarque \& Zinn(1990)]{ldz} Lee, Y.W., Demarque, P., Zinn, R.,
         1990, \apj, 350, 155	 
\bibitem[Lee, Freedman \& Madore(1993)]{lfm93} Lee, M.G., Freedman, W.L, 
         Madore, B.F., 1993, ApJ, 417, 553 (LFM93)
\bibitem[Lub (2002)]{lub} Lub, J., 2002, in $\omega$ Centauri: A Unique
	 Window into Astrophysics, F. van Leeuwen, J. Hughes and G. Piotto Eds.,
	 S. Francisco, ASP, ASP Conf. Series, 265, 95	 
\bibitem[Madore \& Freedman(1995)]{mf95} Madore, B.F, Freedman, W.L., 1995,
         AJ, 109, 1645 (MF95)
\bibitem[Madore \& Freedman(1998)]{mf98} Madore, B.F, Freedman, W.L., 1998,
        in Stellar Astrophysics for the Local Group,      
        A. Aparicio, A. Herrero and F. Sanchez Eds.,
        Cambridge: Cambridge University Press, p. 305
\bibitem[Ma\'iz-Apell\'aniz, Cieza \& MacKenty (2002)]{apella} 
         Ma\'iz-Apell\'aniz, J., Cieza, L., MacKenty, J.W., 2002, AJ, 123, 1601
\bibitem[Mendez et al.(2002)]{mendez} Mendez, B., Davis, M., Moustakas, J., Newman,
         J., Madore, B.F., Freedman, W.L., 2002, AJ, 124, 213	 
\bibitem[Montegriffo et al.(1995)]{paolo} Montegriffo, P., Ferraro, F.R., Fusi
         Pecci, F., Origlia, L., 1995, MNRAS, 276, 739 	 	
\bibitem[Norris \& Da Costa(1995)]{ndc} Norris, J.E., Da Costa, G.S., 1995,
         ApJ, 447, 680
\bibitem[Norris, Freeman \& Mighell(1996)]{n96} Norris, J.E., Freeman, K.C., 
         Mighell, K.L., 1996, ApJ, 462, 241
\bibitem[Olech et al.(2003)]{olech} Olech A., Kaluzny J., Thompson I.B.,
         Schwarzenberg-Czerny A., MNRAS, 345, 86 
\bibitem[Origlia et al.(2003)]{livia} Origlia L., Ferraro F.R., Bellazzini M.,
         Pancino E., 2003, ApJ, 591, 916	 
\bibitem[Pancino et al.(2000)]{panc} Pancino, E., Ferraro, F.R., Bellazzini, M., 
         Piotto G., Zoccali, M., 2000, ApJ, 534, L83
\bibitem[Pancino et al.(2002)]{panc02} Pancino, E., Pasquini L., Hill V.,
         Ferraro F.R., Bellazzini M., 2002, ApJ, 586, L101 	 
\bibitem[Pancino (2003)]{panct}	Pancino, E., 2003, Ph.D. Thesis, Bologna
         University 
\bibitem[Percival et al. (2002)]{perci} Percival, S.M., Salaris, M., van Wyk, F.,
         Kilkenny, D., 2002, ApJ, 573, 174 (P02)	 
\bibitem[Reid (1998)] Reid, I.N., 1998, AJ, 115, 204 (R98)
\bibitem[Renzini (1998)]{alvio} Renzini, A., 1998, AJ, 115, 2459
\bibitem[Renzini \& Buzzoni (1986)]{rebuz} Renzini, A., Buzzoni, A., 1986, in
         Spectral Evolution of Galaxies, C. Chiosi and A. Renzini Eds.,
	 Dordercht, reidel, 135	 
\bibitem[Renzini \& Fusi Pecci(1988)]{rf88} Renzini, A., Fusi Pecci, F., 1988, 
         \araa, 26, 199
\bibitem[Rey et al.(2000)]{rey} Rey, S-C., Lee, Y.W., Joo, J-M., Walker, A., Baird,
         S., 2000, \aj, 119, 1824
\bibitem[Rieke \& Lebofsky(1985)]{rieke} Rieke, G.H., Lebofsky, M.J., 1985, ApJ,
          288, 618	
\bibitem[Robin et al.(2003)]{robin} Robin A.C., Reyl\'e C., Derri\`ere S.,
         Picaud S., 2003, A\&A, 409, 523	  
\bibitem[Rood \& Crocker(1997)]{rood} Rood R.T., Crocker D.A., 1997,
         http://www.astro.virginia.edu/~rtr/papers/mpsvi97.ps.gz	   
\bibitem[Sakai, Madore \& Freedman(1996)]{smf96} Sakai, S., Madore, B.F, 
         Freedman W.L., 1996, ApJ, 461, 713 (SMF96)
\bibitem[Sakai, Zaritsky \& Kennicutt(2000)]{sakai} Sakai S., Zaritsky D.,
         Kennicut R.C.Jr., 2000, AJ, 119, 1197  
\bibitem[Salaris, Chieffi \& Straniero(1993)]{scs93} Salaris, M., Chieffi, A., 
         Straniero O., 1993, ApJ, 414, 580
\bibitem[Salaris \& Cassisi(1998)]{sc98} Salaris, M., Cassisi, S., 1998,
         MNRAS, 298, 166 (SC98)
\bibitem[Salaris \& Weiss (1998)]{sw98} Salaris, M., Weiss, A. 1998, A\&A, 335, 
         943 (SW98)	
\bibitem[Salaris, Cassisi \& Weiss(2002)]{scw} Salaris, M., Cassisi, S., Weiaa, A.,
         2003, PASP, 114, 375
\bibitem[Salaris(2002)]{sal} Salaris M., 2002, in Observed HR diagrams and
         Stellar Evolution: the interplay between observational constraints and
	 theory, T. Lejeune and J. Fernandes Eds., S. Francisco, ASP,
	 ASP Conf. Series, 274, 50 	 
\bibitem[Sarajedini \& Layden(1995)]{sl95} Sarajedini, A., Layden, A.C., 1995,
         \aj, 109, 1086	 
\bibitem[Schechter, Mateo \& Saha(1993)]{dophot} Schechter, P., Mateo, M., Saha,
         A., 1993, PASP, 105, 1342
\bibitem[Shetrone et al.(2003)]{shet} Shetrone M., Venn K.A., Tolstoy E., Primas
         F., Hill V., Kaufer A., 2003, AJ, 125, 684	 
\bibitem[Smith et al.(2002)]{lmcsmith} Smith V.V., Hinkle K.H., Cunha K., et
         al., 2002, AJ, 124, 3241	 	 	 
\bibitem[Sollima et al.(2003)]{antonio} Sollima, A., Ferraro, F.R., Origlia, L.,
         Pancino, E., Bellazzini, M., 2003, A\&A, submitted 
\bibitem[Stetson (2000)]{stet} Stetson, P., 2000, PASP, 112, 925	 	  
\bibitem[Storm et al(1994)]{v9} Storm, Nordstr\"om, B., Carney, B.W., Andersen, J.,
         1994, A\&A, 291, 121 (S94)	
\bibitem[Suntzeff \& Kraft(1996)]{sk96} Suntzeff, N.B., Kraft, R.P., 1996,
         AJ, 111, 1913
\bibitem[Thompson et al.(2001)]{ogle} Thompson, I.B., Kaluzny, J., Pych, W., 
         Burley G., Krzeminski, W., Paczynski, B., Persson, S.E.,
	 Preston G.W., 2001, AJ, submitted (astro-ph/0012493)
\bibitem[Tosi et al.(2001)]{tosi} Tosi, M., Sabbi, E., Bellazzini, M., Aloisi,
         A., Greggio, L., Leitherer, C., Montegriffo, P., 2001, AJ, 122, 1271
\bibitem[Trager (2003)]{trag} Trager, S., in Origin and Evolution of the
         Elements, A. McWilliam and M. Rauch Eds., Cambridge, Cambridge
	 University Press, Carnegie Obs. Astroph. Series, v. 4, in press
	 (astro-ph/0307069) 
\bibitem[Valenti et al.(2004)]{valenti} Valenti E., Ferraro F.R., Perina S.,
         Origlia L., 2004, A\&A, in press (astro-ph/0401153)	 	 	 
\bibitem[van Leeuwen et al.(2000)]{vanleuw}van Leeuwen, F., Poole, R.S., Reijns,
         R.A., Freeman, K.C., de Zeeuw, P.T., 2000, A\&A, 360, 472
\bibitem[Walker (2003)]{walk} Walker, A.R., 2003, in Stellar Candles for the 
         Extragalactic Distance Scale, D. Alloin and W. Gieren Eds., Springer,
	 Lecture Notes in Physics, in press (astro-ph/0303011) 
\bibitem[Zaritsky(1999)]{zarebv} Zaritsky D., 1999, AJ, 118, 2824
\bibitem[Zinn \& West(1984)]{zw84} Zinn, R., West, M.J., 1984, \apjs, 55, 45
\bibitem[Zoccali et al.(2001)]{manu} Zoccali, M., Renzini, A., Ortolani, S., et al.,
         2001, ApJ, 553, 733 (Z01)

\end{thebibliography}
\end{document}